\documentclass[a4paper,11pt]{article}
\usepackage{pos}

\usepackage{cleveref}
\usepackage{easy-todo}

\usepackage{nicefrac}	
\usepackage{braket}

\graphicspath{{figures/},{inkscape/},{../plots/},{tikz/}}

\usepackage{subcaption}
\usepackage{float}
\usepackage{tikz}

\DeclareCaptionLabelFormat{opening}{#2}
\renewcommand{\l}{\left}

\newcommand{\pdagger}{{\phantom{\dagger}}}

\newcommand{\erwartung}[1]{\ensuremath{\left\langle#1\right\rangle}}

\definecolor{mymagenta}{RGB}{200, 0, 100}
\definecolor{myblue}{RGB}{45, 48, 146}

\usepackage{ifthen} 
\usepackage{intcalc} 
\usepackage[skins]{tcolorbox}
\usepackage{pgfplots} 
\usepackage{grffile}  

\usetikzlibrary{external}

\usetikzlibrary{arrows,backgrounds,fit,calc,patterns,shapes}

\usetikzlibrary{decorations.pathmorphing} 
\usetikzlibrary{intersections} 
\usetikzlibrary{positioning}	

\usetikzlibrary{plotmarks}
\pgfplotsset{compat=newest}
\usetikzlibrary{plotmarks}
\usetikzlibrary{arrows.meta}
\usepgfplotslibrary{patchplots}

\tikzset{every node/.style={sloped,allow upside down},baseline={([yshift=+0ex]current bounding box.center)},inner sep=.3mm,x=1cm,y=1cm}

\def\figHeight{0.3\textwidth}
\def\pepsWidth{4mm}

\def\xDist{13mm} 
\def\yDist{10mm}
\def\yExt{\dimexpr \yDist-\pepsWidth /2 \relax}

\def\singularValueWidth{3mm}
\def\swapWidth{1.5mm}
\def\swapHeight{\swapWidth}
\def\gateWidth{\dimexpr \xDist+\pepsWidth \relax}
\def\gateHeight{\pepsWidth}

\tikzset{every picture/.style={node distance=\yDist and \xDist}}

\definecolor{tensorblue}{rgb}{0.8,0.8,1}
\definecolor{tensorred}{rgb}{1,0.5,0.5}
\definecolor{tensorpurple}{rgb}{1,0.5,1}

\tikzset{tenop/.style={fill=orange!30}}
\tikzset{tenk/.style={fill=green!50!black!50}}
\tikzset{tenleft/.style={fill=yellow!50}}
\tikzset{tenright/.style={fill=violet!50}}
\tikzset{tenlr/.style={fill=tensorpurple}}
\tikzset{teneff/.style={fill=tensorpurple}}
\tikzset{tentrans/.style={fill=black!20}}
\tikzset{tens/.style={fill=black!30}}

\tikzset{tenred/.style={fill=tensorred}}
\tikzset{tengreen/.style={fill=green!50!black!50}}
\tikzset{tenpurple/.style={fill=tensorpurple}}
\tikzset{tengrey/.style={fill=black!20}}
\tikzset{tenwhite/.style={fill=black!0}}
\tikzset{tenorange/.style={fill=orange!30}}
\tikzset{u/.style={fill=blue!20,draw=black}}
\tikzset{w/.style={fill=green!50!black!80,draw=black}}

\tikzset{ket/.style={circle,draw=black,thick,fill=tensorblue,minimum width=\pepsWidth}}
\tikzset{bra/.style={ket,tens,label=center:{$*$}}}
\tikzset{ketNew/.style={ket,tengreen}}
\tikzset{braNew/.style={bra,tengreen}}
\tikzset{nodeInvisible/.style={circle,draw=none,thick,minimum width=\pepsWidth}}
\tikzset{bMPS/.style={ket,tengrey}}
\tikzset{singularValue/.style={midway,circle,draw=black,thick,tenorange,minimum width=\singularValueWidth}}
\tikzset{swap/.style={diamond,draw=black,thick,fill=black,minimum width=\swapWidth,minimum height=\swapHeight}}
\tikzset{gate/.style={rectangle,draw=black,thick,fill=tensorpurple,minimum width=\gateWidth,minimum height=\gateHeight}}
\tikzset{env/.style={bMPS,rectangle,draw=black,thick,minimum height=\gateHeight}}

\tikzset{wiggly/.style={decorate, decoration={snake, segment length=\pepsWidth/2, amplitude=\pepsWidth/8}}}
\tikzset{physical/.style={densely dashed}}
\tikzset{parity/.style={dotted,very thick}}
\tikzset{boundary/.style={very thick}}

\newcommand{\connectD}[2]{(#1) -- (#1|-#2.north)}
\newcommand{\connectU}[2]{(#1) -- (#1|-#2.south)}

\title{The Hubbard model with fermionic tensor networks}

\author*[a,b]{Manuel Schneider}
\author[c,d]{Johann Ostmeyer}
\author[a]{Karl Jansen}
\author[c,e,f]{Thomas Luu}
\author[c,d]{Carsten Urbach}

\affiliation[a]{Zeuthen Particle Physics Theory (ZPPT), DESY,\\
	Platanenallee 6, 15738 Zeuthen, Germany}

\affiliation[b]{Institut für Physik,\\
	Humboldt-Universität zu Berlin, Newtonstraße 15, 12489 Berlin, Germany}

\affiliation[c]{Helmholtz-Institut f\"ur Strahlen- und Kernphysik,\\
	University of Bonn, Nussallee 14-16, 53115 Bonn, Germany}

\affiliation[d]{Bethe Center for Theoretical Physics, University of Bonn,\\
	Nussallee 12, 53115 Bonn, Germany}

\affiliation[e]{Institute for Advanced Simulation, Forschungszentrum J\"{u}lich,\\
	54245 J\"{u}lich Germany}

\affiliation[f]{Institut f\"{u}r Kernphysik, Forschungszentrum J\"{u}lich,\\
	54245 J\"{u}lich Germany}

\emailAdd{manuel.schneider@desy.de}

\abstract{Many electromagnetic properties of graphene can be described by the Hubbard model on a honeycomb lattice. However, this system suffers strongly from the sign problem if a chemical potential is included. Tensor network methods are not affected by this problem. We use the imaginary time evolution of a fermionic projected entangled pair state, which allows to simulate both parity sectors independently. Incorporating the fermionic nature on the level of the tensor network allows to fix the particle number to be either even or odd. This way we can access the states at half filling and with one additional electron. We calculate the energy and other observables of both states, which was not possible before with Monte Carlo methods.}

\FullConference{%
 The 38th International Symposium on Lattice Field Theory, LATTICE2021
  26th-30th July, 2021
  Zoom/Gather@Massachusetts Institute of Technology
}

\begin{document}
\maketitle

\section{Introduction}
In lattice field calculations Quantum Monte Carlo (QMC) methods are well established and a powerful tool to simulate physical systems. However, parts of the phase space are often inaccessible due to the infamous sign problem. In these cases alternative methods have to be established, tested and improved. One such physical system is the Hubbard model with a chemical potential, which often serves as a test ground for new methods.

The Hubbard model on a honeycomb lattice, as studied here, is a model for graphene. The material graphene has gained a lot of attention in recent years with a vast range of applications~\cite{Novoselov2007}. The model undergoes a first order phase transition from a semi-metallic to a Mott-insulating phase \cite{Ostmeyer:2020uov}. The semi-metallic behavior of graphene, for example, was used to build transistors with a much higher switching rate than silicon-based transistors~\cite{graphene_semiconductor}.

The Hubbard model is well studied theoretically as well. The Hamiltonian reads as

\begin{align}
	H&=-\kappa \sum_{\erwartung{x,y},s}c^\dagger_{x,s}c^\pdagger_{y,s}+\frac{U}{2}\sum_{x}q_x^2 + B \sum_{x}\left(c_{x,\uparrow}^\dagger c_{x,\uparrow}^\pdagger-c_{x,\downarrow}^\dagger c_{x,\downarrow}^\pdagger \right)
	+\mu \sum_{x,s}\left(c_{x,s}^\dagger c_{x,s}^\pdagger-\frac12\right) .
	\label{eq:hubbard_hamilton}
\end{align}
The first term is called the hopping term which allows the electrons to jump to a nearest neighbor. The corresponding coupling strength is $\kappa$. The sum runs over all nearest neighbors $\erwartung{x,y}$ on the honeycomb lattice, whereas $s\in\left\{\uparrow,\,\downarrow\right\}$ denotes the two possibilities for the spin of an electron. $c^\dagger_{x,s}$ ($c^\pdagger_{x,s}$) is a creation (annihilation) operator of an electron at position $x$ with spin $s$. The second term is the onsite interaction with a strength $U$ connected to the local charge $q_x=c^\dagger_{x,\uparrow}c^\pdagger_{x,\uparrow}+c^\dagger_{x,\downarrow}c^\pdagger_{x,\downarrow}-1$. An external magnetic field of strength $B$ changes the energy according to the magnetization of the system. The last term couples the chemical potential $\mu$ to the total electron number.

The most established method to study the system is QMC. Many results are available, especially on the nature of the phase transition~\cite{Ostmeyer:2020uov,Assaad:2013xua,Buividovich:2018crq}. However, the stochastic Monte Carlo method suffers from the sign problem if the chemical potential is included. Only the case of half filling can directly be simulated, where the chemical potential vanishes and the number of electrons is half its maximum. Other methods than QMC are needed to explore the phase space including the chemical potential.

Tensor network based approaches are immune to the sign problem and are therefore a perfect choice. We use a fermionic projected entangled pairs state (PEPS)~\cite{tensor_fermions,iPEPS_introduction} as an ansatz to find the ground state. Moreover, we can restrict the state to a specific parity sector. This allows us to compute the properties of the state with one additional electron or hole compared to the ground state. More details of our methods and further results can be found in \cite{Schneider2021}. Here, we give a short introduction regarding the method and summarize the most important results. In addition to \cite{Schneider2021} we also present the dependence of the measured particle number on the chemical potential for a small system.

\section{Fermionic PEPS}
To implement the fermionic anti-commutation relations on the level of the PEPS ansatz, we closely follow the proposal of \cite{tensor_fermions}. The main idea is to define the parity on all links of the tensor network and to introduce swap gates that ensure the correct sign flips when fermions are exchanged. Additionally, the PEPS consists of only even parity tensors. This allows for a simple graphical representation of the fermionic PEPS, see~\cref{fig:peps_ket}, where tensors are denoted as nodes, indices correspond to links and swap gates, represented as diamonds, are placed on all line crossings in the two-dimensional graphical representation. Lines can then be moved freely, even across nodes, if new line crossings get equipped with swap gates.

\begin{figure}[H]
	\begin{minipage}[t]{0.45\textwidth}
		\centering
		\begin{tikzpicture}
	\def\Lx{3}
	\def\Ly{4}
	\def\xStep{\xDist}
	\def\yStep{-\yDist}
	\def\physLength{-\yExt}
	\def\xShift{\xStep / \Ly} 
	\node (origin) at (- \xShift*3, \yStep) {};
	\draw[-latex] (origin.center) -- +(\xStep, 0) node[midway,label=below:x] {};
	\draw[-latex] (origin.center) -- +(-\xShift, \yStep) node[midway,label=above:y] {};
	\foreach \x in {1,...,\Lx}
		\foreach \y in {1,...,\Ly}
			{
				\node[ket] (ten_\x_\y) at (\xStep*\x - \xShift * \y, \yStep*\y) {};
				\node (physEnd_\x_\y) at (\xStep*\x - \xShift * \y, {\yStep*\Ly + \physLength}) {};
			}
		\draw[parity] (ten_1_\Ly) -- (-\xShift,{\yStep*\Ly + \physLength});
		\foreach \x in {1,...,\Lx}
			\foreach \y in {1,...,\Ly}
			{
				\draw[physical,name path={phys_\x_\y}] (ten_\x_\y) -- (physEnd_\x_\y);
			}
		\foreach \x in {2,...,\Lx}
			\foreach \y in {1,...,\Ly}
			{
				\ifthenelse{\intcalcMod{\x+\y}{2}=0}{
				}{
					\pgfmathtruncatemacro\lastx{\x-1}
					\pgfmathtruncatemacro\lasty{\y-1}
					\draw[name path={xlink_\lastx_\y}] (ten_\lastx_\y) -- (ten_\x_\y);
					\ifthenelse{\y>1}{ 
						\foreach \physy in {1,...,\lasty}
						{
							\path [name intersections={of={xlink_\lastx_\y} and {phys_\lastx_\physy},by={swap_\lastx_\y_\physy}}];
							\node[swap]  at ({swap_\lastx_\y_\physy}) {};
						}
					}
				}
			}
		\foreach \x in {1,...,\Lx}
			\foreach \y in {2,...,\Ly}
			{
				\pgfmathtruncatemacro\lasty{\y-1}
				\draw (ten_\x_\lasty) -- (ten_\x_\y);
			}
\end{tikzpicture}
		\caption{Fermionic PEPS on a honeycomb lattice. Circles: PEPS tensors; solid lines: internal indices; dashed lines: physical indices; dotted line: parity index; diamonds: swap gates; see~\cite{Schneider2021} for details.}
		\label{fig:peps_ket}
	\end{minipage}
	\hfill
	\begin{minipage}[t]{0.45\textwidth}
		\centering
		\begin{tikzpicture}
	\def\Lx{5}
	\def\xDistLocal{\xDist*2/3}
	\def\xShift{\xDist*0.9/2} 
	\def\yShift{-\xShift} 
	\def\yShiftBra{-\xShift*2} 
	\def\bMPSyShift{\yShift/5}
	\node[ket] (ketL) at (0, 0) {};
	\node[bra] (braL) at ([shift=({\xShift,\yShiftBra})]ketL.center) {};
	\node[bMPS] (bMPSxLU) at ([shift=({\xShift/2,-\yShift})]ketL.center) {};
	\draw (ketL) .. controls ([shift=({-\xShift/2,\bMPSyShift})]bMPSxLU.center) .. (bMPSxLU);
	\draw[name path={bMPSxlinkLU}] (braL) .. controls ([shift=({\xShift/2,\bMPSyShift})]bMPSxLU.center) .. (bMPSxLU);
	\node[bMPS] (bMPSxLD) at ([shift=({-\xShift/2,\yShift})]braL.center) {};
	\draw[name path={bMPSxlinkLD}] (ketL) .. controls ([shift=({-\xShift/2,-\bMPSyShift})]bMPSxLD.center) .. (bMPSxLD);
	\draw (braL) .. controls ([shift=({\xShift/2,-\bMPSyShift})]bMPSxLD.center) .. (bMPSxLD);
	\node[bMPS] (bMPSyL) at ([shift=({\yShift,\yShiftBra/2})]ketL.center) {};
	\draw (ketL) .. controls ([shift=({0,-\yShiftBra/2})]bMPSyL.center) and ([shift=({-\bMPSyShift,\xShift/2})]bMPSyL.center) .. (bMPSyL);
	\draw[name path={bMPSylinkL}] (braL) .. controls ([shift=({0,\yShiftBra/2})]bMPSyL.center) and ([shift=({-\bMPSyShift,-\xShift/2})]bMPSyL.center) .. (bMPSyL);
	\path[name intersections={of={bMPSxlinkLD} and {bMPSylinkL},by={swapLD}}];
	\node[swap]  at (swapLD) {};
	\node[ket] (triadLU) at ([shift=({\xDistLocal+\xShift,0})]ketL.center) {};
	\draw[wiggly,name path={wigglyLU}] (ketL) -- (triadLU);
	\path[name intersections={of={bMPSxlinkLU} and {wigglyLU},by={swapLU}}];
	\node[swap]  at (swapLU) {};
	\node[bra] (triadLD) at ([shift=({0,\yShiftBra})]triadLU.center) {};
	\draw[wiggly] (braL) -- (triadLD);
	\node[ket] (triadRU) at ([shift=({\xDistLocal,0})]triadLU.center) {};
	\draw (triadLU) -- (triadRU) node[midway,label={above:$D$}] (middle) {};
	\node[bra] (triadRD) at ([shift=({\xDistLocal,0})]triadLD.center) {};
	\draw (triadLD) -- (triadRD) node[midway,label={below:$D$}] {};;
	\node[ket] (ketR) at ([shift=({\xDistLocal,0})]triadRU.center) {};
	\draw[wiggly] (triadRU) -- (ketR);
	\node[bra] (braR) at ([shift=({\xShift,\yShiftBra})]ketR.center) {};
	\draw[wiggly,name path={wigglyRD}] (triadRD) -- (braR);
	\node[bMPS] (bMPSxRU) at ([shift=({\xShift/2,-\yShift})]ketR.center) {};
	\draw (ketR) .. controls ([shift=({-\xShift/2,\bMPSyShift})]bMPSxRU.center) .. (bMPSxRU);
	\draw[name path={bMPSxlinkRU}] (braR) .. controls ([shift=({\xShift/2,\bMPSyShift})]bMPSxRU.center) .. (bMPSxRU);
	\node[bMPS] (bMPSxRD) at ([shift=({-\xShift/2,\yShift})]braR.center) {};
	\draw[name path={bMPSxlinkRD}] (ketR) .. controls ([shift=({-\xShift/2,-\bMPSyShift})]bMPSxRD.center) .. (bMPSxRD);
	\draw (braR) .. controls ([shift=({\xShift/2,-\bMPSyShift})]bMPSxRD.center) .. (bMPSxRD);
	\path[name intersections={of={wigglyRD} and {bMPSxlinkRD},by={swapRD}}];
	\node[swap]  at (swapRD) {};
	\node[bMPS] (bMPSyR) at ([shift=({\xShift,-\yShiftBra/2})]braR.center) {};
	\draw[name path={bMPSylinkR}] (ketR) .. controls ([shift=({0,-\yShiftBra/2})]bMPSyR.center) and ([shift=({\bMPSyShift,\xShift/2})]bMPSyR.center) .. (bMPSyR);
	\draw (braR) .. controls ([shift=({0,\yShiftBra/2})]bMPSyR.center) and ([shift=({\bMPSyShift,-\xShift/2})]bMPSyR.center) .. (bMPSyR);
	\path[name intersections={of={bMPSxlinkRU} and {bMPSylinkR},by={swapRU}}];
	\node[swap]  at (swapRU) {};
	\draw[boundary] (bMPSxLU) -- (bMPSxRU) node[midway,label={above:$\chi$}] {};
	\draw[boundary] (bMPSxRU) .. controls ([shift=({\xShift+\xShift/2,0})]bMPSxRU.center) .. (bMPSyR) node[midway,label={above:$\chi$}] {};;
	\draw[boundary] (bMPSxRD) .. controls ([shift=({\xShift+\xShift/2,0})]bMPSxRD.center) .. (bMPSyR) node[midway,label={below:$\chi$}] {};;
	\draw[boundary] (bMPSxLD) -- (bMPSxRD) node[midway,label={below:$\chi$}] {};;
	\draw[boundary] (bMPSxLD) .. controls ([shift=({-\xShift-\xShift/2,0})]bMPSxLD.center) .. (bMPSyL) node[midway,label={above:$\chi$}] {};;
	\draw[boundary] (bMPSxLU) .. controls ([shift=({-\xShift-\xShift/2,0})]bMPSxLU.center) .. (bMPSyL) node[midway,label={below:$\chi$}] {};;
	\node[gate, minimum width = \xDistLocal+\pepsWidth] (operator) at ([shift=({0,\yShiftBra/2})]middle) {$\hat{o}$};
	\draw[physical] \connectD{triadLU}{operator};
	\draw[physical] \connectD{triadRU}{operator};
	\draw[physical] \connectU{triadLD}{operator};
	\draw[physical] \connectU{triadRD}{operator};
\end{tikzpicture}
		\caption{Calculation of the expectation value of a two-site operator $\hat{o}$. Gray circles: boundary MPS tensors; thick solid lines: boundary MPS indices of dimension $\chi$; violet circles and wiggly lines: PEPS tensors split by a QR decomposition; gray circles with stars: conjugated PEPS tensors with certain swap gates applied; see~\cite{Schneider2021} for details.}
		\label{fig:expectation_value}
	\end{minipage}
\end{figure}


We use a PEPS on a honeycomb lattice as depicted in \cref{fig:peps_ket} as an ansatz for the states. Since it consists of tensors of even parity, it can only be used for states with overall even parity. Therefore, we include an additional parity link on one of the tensors, see the dotted line in \cref{fig:peps_ket}. This index has only one value of either even or odd parity, which defines the overall parity of the state. In this manner we are able to simulate the even- and odd-parity sectors independently, corresponding to the ground state and a single particle excitation with one additional electron or hole.

\section{Algorithm and Improvements}
We use an imaginary time evolution of the Trotter-decomposed Hamiltonian~\cite{SuzukiTrotter_MPS,iPEPS_introduction} in order to find the ground state in each parity sector. The bond dimension grows when a local gate is applied to neighboring tensors during the evolution. Using the simple update routine, we truncate the singular values to keep the bond dimension $D$ bounded~\cite{tensor_fermions,Lubasch_algorithms,iPEPS_introduction}. While decreasing the imaginary time step in each update step, the system eventually converges to the ground state. An optimization of the imaginary time evolution step sizes $\delta t$ and an extrapolation in $\delta t$ allows for an efficient usage of resources. For the optimized evolution we introduce a way to estimate the ground state energy without contracting the whole network~\cite{Schneider2021}. 

Finally, expectation values like the energy or the particle number can be measured. An exact calculation of these observables requires a sum over all indices of the tensor network, which scales exponentially in the length of the system and is therefore not applicable for larger volumes. Instead, we use the boundary matrix product state~(MPS) method \cite{Lubasch_algorithms} to approximate parts of the network. The calculation of an expectation value of a two site operator is shown in \cref{fig:expectation_value}. The operator acts on two neighboring tensors and their conjugates. The rest of the network is approximated by a boundary MPS with a bond dimension of $\chi$, shown as gray circles without stars.

\begin{figure}[ht]
	\centering
%
%
\definecolor{mycolor1}{rgb}{0.898,0.122,0.059}%
\definecolor{mycolor2}{rgb}{0.58,0,0.831}%
\definecolor{mycolor3}{rgb}{0.00000,0.62,0.451}%
\definecolor{mycolor4}{rgb}{0.341,0.71,0.91}%
\definecolor{mycolor5}{rgb}{0.898,0.62,0}%
\begin{tikzpicture}

\begin{axis}[%
width=.75\linewidth,
height=0.6\textwidth,
unbounded coords=jump,
xmin=1.5,
xmax=180,
xlabel style={font=\color{white!15!black}},
xlabel={$\chi$},
xtick={10, 100, 144},
xticklabels={$10^1$, $10^2$, $144^{\phantom{1}}$},
xmode=log,
ymode=log,
ymin=1e-06,
ymax=2e-01,
yminorticks=true,
ylabel style={font=\color{white!15!black},{rotate=-90}},
ylabel={$\left|\nicefrac{\Braket{O\left(\chi\right)}}{\Braket{O\left(144\right)}}-1 \right|$},
axis background/.style={fill=white},
title style={font=\bfseries},
xmajorgrids,
ymajorgrids,
legend style={legend cell align=left, align=left, draw=white!15!black, 
	legend pos=outer north east}
]
\addplot [color=black, only marks, mark=x, mark options={solid, black, scale=1.5, thick}]
	table[row sep=crcr]{%
		2	2.622291e-02\\
		4	1.361124e-02\\
		8	6.714825e-03\\
		16	2.286470e-03\\
		32	4.737575e-04\\
		64	5.666863e-05\\
		81	1.793656e-05\\
		100	5.132277e-06\\
		144	0\\
	};
\addlegendentry{Energy $H$}
\addplot [color=mycolor2, only marks, mark=o, mark options={solid, mycolor2, scale=1.5, thick}]
	table[row sep=crcr]{%
		2	6.021158e-02\\
		4	1.406397e-02\\
		8	1.895159e-02\\
		16	6.193285e-04\\
		32	8.671524e-04\\
		64	1.781735e-04\\
		81	6.983127e-05\\
		100	7.214596e-05\\
		144	0\\
	};
\addlegendentry{Magnetization $M$}
\addplot [color=mycolor4, only marks, mark=triangle, mark options={rotate=180, solid, mycolor4, scale=1.5, thick}]
	table[row sep=crcr]{%
		2	5.831624e-05\\
		4	1.296688e-04\\
		8	5.619606e-04\\
		16	2.843423e-06\\
		32	4.763113e-05\\
		64	4.132079e-06\\
		81	3.177223e-06\\
		100	3.601051e-06\\
		144	0\\
	};
\addlegendentry{Particle number $n$}
\addplot [color=mycolor5, only marks, mark=+, mark options={solid, mycolor5, scale=1.5, thick}]
	table[row sep=crcr]{%
		2	4.141434e-03\\
		4	9.208664e-03\\
		8	3.990866e-02\\
		16	2.019309e-04\\
		32	3.382612e-03\\
		64	2.934472e-04\\
		81	2.256363e-04\\
		100	2.557353e-04\\
		144	0\\
	};
\addlegendentry{Hole number $V-n$}
\addplot [color=mycolor1, only marks, mark=square, mark options={solid, mycolor1, scale=1.5, thick}]
table[row sep=crcr]{%
	2	1.113176e-01\\
	4	1.134130e-01\\
	8	1.553331e-02\\
	16	9.732163e-03\\
	32	8.573809e-03\\
	64	1.892462e-03\\
	81	1.004796e-03\\
	100	2.651277e-04\\
	144	1.134541e-04\\
};
\addlegendentry{Norm deviation $\Delta I$\\(absolute values)}
\end{axis}
\end{tikzpicture}%
	\caption{Dependence of different operators $\Braket{O}$ on the boundary MPS bond dimension $\chi$ for the ground state in the odd parity sector. Shown are the relative errors compared to results at $\chi = 144$ and absolute values for the standard deviation of the norm. Honeycomb Hubbard model with $D=12$, lattice size $V=12\times6$, $\kappa=1$, $U=2$, $\mu=0.1$, $B=0.01$. See~\cite{Schneider2021,raw_data_arXiv} for details.}
	\label{fig:boundaryMPS}
\end{figure}
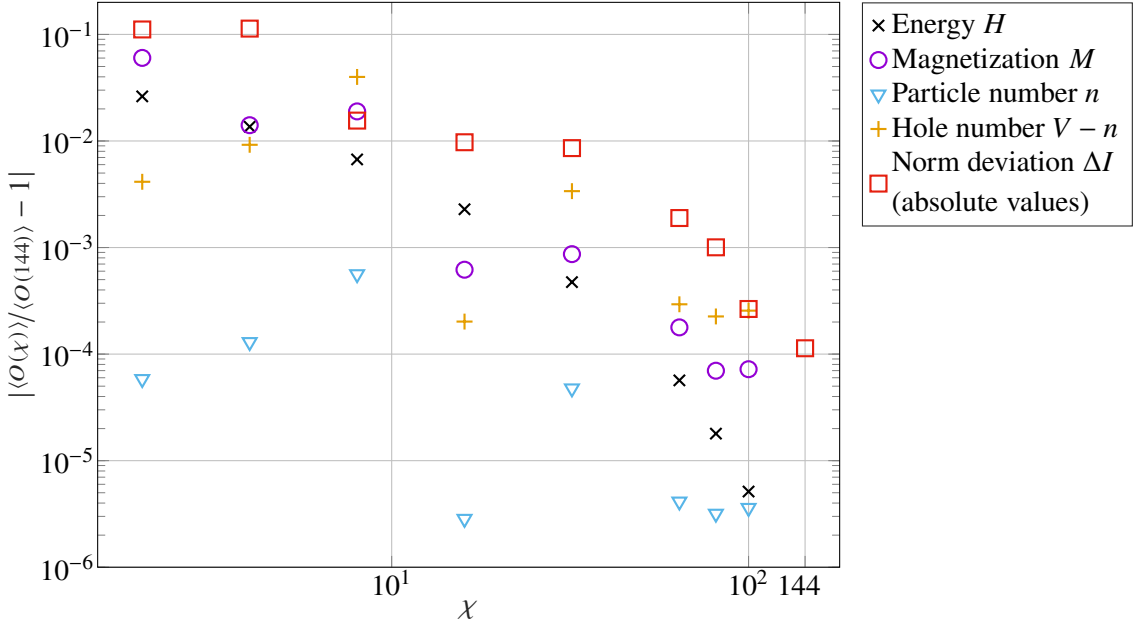

In previous simulations $\chi \ge D^2$ was typically chosen~\cite{tensor_fermions,Lubasch_algorithms}. We find that a much smaller boundary MPS bond dimension is sufficient to gain an acceptable accuracy. In~\cref{fig:boundaryMPS} the relative errors of several observables measured in a ground state are shown when different boundary MPS bond dimensions are used. The results are compared to $\chi = D^2 = 144$. A sub percent accuracy is already obtained with $\chi = 16$ for these observables. In further simulations we therefore calculate expectation values with $\chi = 2D$ and $\chi = 3D$ and find no significant deviations in most cases. This allows the use of larger bond dimensions $D$ to gain a better overall accuracy.

The standard deviation of the norm $\Delta I$ evaluated at different lattice sites~\cite{Schneider2021} gives a good approximation of the relative errors introduced by a finite $\chi$, as seen in \cref{fig:boundaryMPS}.

The algorithm is less accurate or can even become unstable if the ground state is degenerate. To solve this problem in the case of $\mu = B = 0$ we introduce a small chemical potential $\mu$ and magnetic field $B$ to break these degeneracies. These small, perturbative effects can afterwards be removed~\cite{Schneider2021}. In cases where the energy gap $\Delta_{ex}$ in one sector is small compared to the single particle excitation gap $\Delta_{sp}$, however, simple update faces instabilities that can not be cured so easily. Using a non-local update procedure like full update~\cite{tensor_fermions,iPEPS_introduction} can possibly help to increase the range of parameters where stable simulations are possible.

\section{Results}
We study the $3 \times 4$ lattice with open boundary conditions and calculate the energy and particle number for different chemical potentials. For this moderate system size the results can be compared to those of an exact diagonalization. \Cref{fig:energy_sectors} shows the energies of the even- and odd- parity sectors. They converge to the correct values if the bond dimension increases. By subtracting the energies of the even- and odd-parity ground states for a given value of $\mu$, we extract the single particle excitation gap $\Delta_{sp} = \left| E_\text{even} - E_\text{odd} \right|$ as shown in \cref{fig:energy_gap}. Finally, \cref{fig:particle_number} shows the measured particle number $n$.

\begin{figure}[ht]
	\centering
	\begin{subfigure}[b]{\textwidth}
		\centering
%
%
\definecolor{mycolor1}{rgb}{0.89800,0.12200,0.05900}%
\definecolor{mycolor2}{rgb}{0.58000,0.00000,0.83100}%
\definecolor{mycolor3}{rgb}{0.00000,0.62000,0.45100}%
\definecolor{mycolor4}{rgb}{0.34100,0.71000,0.91000}%
\definecolor{mycolor5}{rgb}{0.89800,0.62000,0.00000}%
\definecolor{mycolor6}{rgb}{0.00000,0.45100,0.70200}%
\begin{tikzpicture}

\begin{axis}[%
width=0.765\textwidth,
height=\figHeight,
at={(0\textwidth,0\textwidth)},
scale only axis,
xmin=0,
xmax=1.05,
xlabel style={font=\color{white!15!black}},
xlabel={$\mu$},
ymin=-11.8,
ymax=-10,
ylabel style={font=\color{white!15!black}},
ylabel={$E$},
axis background/.style={fill=white},
axis x line*=bottom,
axis y line*=left,
xmajorgrids,
ymajorgrids,
legend style={at={(1.03,1)}, anchor=north west, legend cell align=left, align=left, draw=white!15!black},
ylabel style={{rotate=-90}}, scaled ticks=false, tick label style={/pgf/number format/fixed},
ytick={-10, -10.5, -11, -11.5},
yticklabels={-10, -10.5, -11, -11.5}
]
\addplot [color=black, line width=1.5]
  table[row sep=crcr]{%
0	-10.5298153648625\\
0.43181417308775	-10.5298153648625\\
1.05	-11.766187018687\\
};
\addlegendentry{exact}

\addplot [color=mycolor1, only marks, mark=o, mark options={solid, mycolor1, scale=1.5, thick}]
  table[row sep=crcr]{%
0	-10.5001002528863\\
0	-10.5062386305375\\
0.1	-10.5000464909614\\
0.1	-10.5062446183013\\
1	-11.6409648893093\\
1	-11.6408502806912\\
0.2	-10.5001791334991\\
0.2	-10.5059192434457\\
0.3	-10.5001715890363\\
0.3	-10.5059675616637\\
0.4	-10.5004520076535\\
0.4	-10.5060312311005\\
0.5	-10.6547285559836\\
0.5	-10.6554096416727\\
0.6	-10.8551357574856\\
0.6	-10.8558379259004\\
0.7	-11.0551088764273\\
0.7	-11.0561361246717\\
0.8	-11.2556202674509\\
0.8	-11.2565295229987\\
0.9	-11.4566428355544\\
0.9	-11.4573701034263\\
};
\addlegendentry{D=20}

\addplot [color=mycolor2, only marks, mark=triangle, mark options={solid, mycolor2, scale=1.5, thick}]
  table[row sep=crcr]{%
0	-10.4943266954245\\
0	-10.4969633620988\\
0.1	-10.4930052329351\\
0.1	-10.4954717517571\\
1	-11.6378704468114\\
1	-11.6386168287065\\
0.2	-10.4928296319942\\
0.2	-10.4954698817055\\
0.3	-10.4935976264209\\
0.3	-10.4958392614775\\
0.4	-10.4932708470626\\
0.4	-10.4953587970669\\
0.5	-10.6435633185814\\
0.5	-10.6458718486759\\
0.6	-10.843818847898\\
0.6	-10.8472804231615\\
0.7	-11.0441568108442\\
0.7	-11.0473404577313\\
0.8	-11.2452408459727\\
0.8	-11.248012800336\\
0.9	-11.4502824174868\\
0.9	-11.4502588858753\\
};
\addlegendentry{D=18}

\addplot [color=mycolor3, only marks, mark=triangle, mark options={solid, rotate=180, mycolor3, scale=1.5, thick}]
  table[row sep=crcr]{%
0	-10.4612448027389\\
0	-10.4622602194599\\
0.1	-10.4612831693327\\
0.1	-10.4622811794776\\
1	-11.6067527194471\\
1	-11.6101873894787\\
0.2	-10.4614422397529\\
0.2	-10.4623877506788\\
0.3	-10.461476213384\\
0.3	-10.4624232283983\\
0.4	-10.461584636767\\
0.4	-10.4625047455177\\
0.5	-10.4623408877235\\
0.5	-10.4631286218273\\
0.6	-10.8243181819159\\
0.6	-10.8260542331005\\
0.7	-11.0162490430863\\
0.7	-11.0211100437376\\
0.8	-11.2226604639981\\
0.8	-11.2257440007429\\
0.9	-11.4297577144851\\
0.9	-11.4307481179588\\
};
\addlegendentry{D=14}

\addplot [color=mycolor4, only marks, mark=+, mark options={solid, mycolor4, scale=1.5, thick}]
  table[row sep=crcr]{%
0	-10.3872319096744\\
0	-10.3885370432664\\
0.1	-10.387160537593\\
0.1	-10.3885229392545\\
1	-11.4976868122309\\
1	-11.493036193671\\
0.2	-10.3861535252034\\
0.2	-10.387476796155\\
0.3	-10.3873406834508\\
0.3	-10.3886520407535\\
0.4	-10.3877628351819\\
0.4	-10.3890532839191\\
0.5	-10.3879835019855\\
0.5	-10.3889346276579\\
0.6	-10.764526398719\\
0.6	-10.7652259431016\\
0.7	-10.9656017602587\\
0.7	-10.9660384941182\\
0.8	-11.1646365433271\\
0.8	-11.1645589952197\\
0.9	-11.3655696033643\\
0.9	-11.3650353808809\\
};
\addlegendentry{D=10}

\addplot [color=mycolor5, only marks, mark=x, mark options={solid, mycolor5, scale=1.5, thick}]
  table[row sep=crcr]{%
0	-10.3441566281981\\
0	-10.3441412759339\\
0.1	-10.3439982238301\\
0.1	-10.3441401737635\\
1	-11.3042916250715\\
1	-11.3057953121656\\
0.2	-10.3440499393303\\
0.2	-10.3441367992347\\
0.3	-10.3440967193712\\
0.3	-10.3441293872783\\
0.4	-10.3441650032126\\
0.4	-10.3441161962261\\
0.5	-10.3441253266366\\
0.5	-10.3441053354384\\
0.6	-10.3440530343431\\
0.6	-10.3441012189147\\
0.7	-10.8225361648016\\
0.7	-10.8223739738321\\
0.8	-11.0384752817504\\
0.8	-11.0380720698294\\
0.9	-11.2480868892368\\
0.9	-11.2474412388606\\
};
\addlegendentry{D= 8}

\addplot [color=mycolor6, only marks, mark=diamond, mark options={solid, mycolor6, scale=1.5, thick}]
  table[row sep=crcr]{%
0	-10.0765143555121\\
0	-10.0776524607606\\
0.1	-10.0765192867472\\
0.1	-10.0777966056128\\
1	-10.978134066071\\
1	-10.978409130489\\
0.2	-10.0771080580798\\
0.2	-10.0784221600203\\
0.3	-10.0784545400239\\
0.3	-10.0795531710436\\
0.4	-10.0802211356126\\
0.4	-10.0813841238161\\
0.5	-10.083416775022\\
0.5	-10.0843492067681\\
0.6	-10.2116265998302\\
0.6	-10.2121126592747\\
0.7	-10.4355980491288\\
0.7	-10.4360332957001\\
0.8	-10.6343603410158\\
0.8	-10.6347449833018\\
0.9	-10.8353424207126\\
0.9	-10.8356210467582\\
};
\addlegendentry{D= 5}

\end{axis}
\end{tikzpicture}%
		\caption{Even parity}
		\label{fig:energy_even}
	\end{subfigure}
	\begin{subfigure}[b]{\textwidth}
		\centering
%
%
\definecolor{mycolor1}{rgb}{0.89800,0.12200,0.05900}%
\definecolor{mycolor2}{rgb}{0.58000,0.00000,0.83100}%
\definecolor{mycolor3}{rgb}{0.00000,0.62000,0.45100}%
\definecolor{mycolor4}{rgb}{0.34100,0.71000,0.91000}%
\definecolor{mycolor5}{rgb}{0.89800,0.62000,0.00000}%
\definecolor{mycolor6}{rgb}{0.00000,0.45100,0.70200}%
\begin{tikzpicture}

\begin{axis}[%
width=0.765\textwidth,
height=\figHeight,
scale only axis,
xmin=0,
xmax=1.05,
xlabel style={font=\color{white!15!black}},
xlabel={$\mu$},
ymin=-11.7,
ymax=-8.5,
ylabel style={font=\color{white!15!black}},
ylabel={$E$},
axis background/.style={fill=white},
axis x line*=bottom,
axis y line*=left,
xmajorgrids,
ymajorgrids,
legend style={at={(1.03,1)}, anchor=north west, legend cell align=left, align=left, draw=white!15!black},
ylabel style={{rotate=-90}}, scaled ticks=false, tick label style={/pgf/number format/fixed},
ytick={-11, -11, -10,-9},
yticklabels={\phantom{-11.5}, -11, -10, -9}
]
\addplot [color=black, line width=1.5]
  table[row sep=crcr]{%
0	-10.204332431282\\
0.86203046922805	-11.0663629005101\\
1.05	-11.6302714928259\\
};
\addlegendentry{exact}

\addplot [color=mycolor1, only marks, mark=o, mark options={solid, mycolor1, scale=1.5, thick}]
  table[row sep=crcr]{%
0	-10.6077\\
0	-8.71885\\
0.1	-10.2843\\
0.1	-10.2862\\
1	-11.4634\\
1	-11.4607\\
0.2	-10.3846\\
0.2	-10.3863\\
0.3	-10.4853\\
0.3	-10.4874\\
0.4	-10.586\\
0.4	-10.5876\\
0.5	-10.687\\
0.5	-10.6886\\
0.6	-10.7871\\
0.6	-10.7881\\
0.7	-10.8846\\
0.7	-10.8852\\
0.8	-10.9186\\
0.8	-10.9214\\
0.9	-11.1631\\
0.9	-11.161\\
};
\addlegendentry{D=20}

\addplot [color=mycolor2, only marks, mark=triangle, mark options={solid, mycolor2, scale=1.5, thick}]
  table[row sep=crcr]{%
0	-12.4363\\
0	-11.839\\
0.1	-10.2811\\
0.1	-10.2826\\
1	-11.455\\
1	-11.4544\\
0.2	-10.382\\
0.2	-10.3828\\
0.3	-10.4816\\
0.3	-10.4839\\
0.4	-10.5824\\
0.4	-10.5838\\
0.5	-10.6824\\
0.5	-10.6841\\
0.6	-10.7825\\
0.6	-10.7838\\
0.7	-10.8814\\
0.7	-10.8824\\
0.8	-10.9104\\
0.8	-10.9111\\
0.9	-11.1524\\
0.9	-11.1532\\
};
\addlegendentry{D=18}

\addplot [color=mycolor3, only marks, mark=triangle, mark options={solid, rotate=180, mycolor3, scale=1.5, thick}]
  table[row sep=crcr]{%
0	-10.1613\\
0	-10.1644\\
0.1	-10.2618\\
0.1	-10.265\\
1	-11.4358\\
1	-11.4358\\
0.2	-10.3624\\
0.2	-10.3646\\
0.3	-10.4627\\
0.3	-10.465\\
0.4	-10.5629\\
0.4	-10.5651\\
0.5	-10.6628\\
0.5	-10.6651\\
0.6	-10.7632\\
0.6	-10.7653\\
0.7	-10.8358\\
0.7	-10.8372\\
0.8	-10.8722\\
0.8	-10.8751\\
0.9	-11.1365\\
0.9	-11.136\\
};
\addlegendentry{D=14}

\addplot [color=mycolor4, only marks, mark=+, mark options={solid, mycolor4, scale=1.5, thick}]
  table[row sep=crcr]{%
0	-10.0844\\
0	-10.0867\\
0.1	-10.1846\\
0.1	-10.1869\\
1	-11.3939\\
1	-11.3939\\
0.2	-10.2841\\
0.2	-10.2866\\
0.3	-10.3847\\
0.3	-10.3868\\
0.4	-10.4924\\
0.4	-10.4948\\
0.5	-10.5928\\
0.5	-10.5949\\
0.6	-10.6944\\
0.6	-10.6965\\
0.7	-10.7968\\
0.7	-10.7984\\
0.8	-10.8972\\
0.8	-10.8988\\
0.9	-11.0919\\
0.9	-11.0914\\
};
\addlegendentry{D=10}

\addplot [color=mycolor5, only marks, mark=x, mark options={solid, mycolor5, scale=1.5, thick}]
  table[row sep=crcr]{%
0	-9.98125\\
0	-9.98288\\
0.1	-10.0882\\
0.1	-10.0893\\
1	-11.3046\\
1	-11.306\\
0.2	-10.1888\\
0.2	-10.1898\\
0.3	-10.2911\\
0.3	-10.2921\\
0.4	-10.3995\\
0.4	-10.4004\\
0.5	-10.5\\
0.5	-10.5008\\
0.6	-10.6004\\
0.6	-10.6011\\
0.7	-10.7007\\
0.7	-10.7012\\
0.8	-10.7051\\
0.8	-10.7064\\
0.9	-11.0048\\
0.9	-10.9693\\
};
\addlegendentry{D= 8}

\addplot [color=mycolor6, only marks, mark=diamond, mark options={solid, mycolor6}]
  table[row sep=crcr]{%
0	-9.62086\\
0	-9.62043\\
0.1	-9.72134\\
0.1	-9.72091\\
1	-11.0206\\
1	-11.0208\\
0.2	-9.82184\\
0.2	-9.82138\\
0.3	-9.92231\\
0.3	-9.92185\\
0.4	-10.0227\\
0.4	-10.0223\\
0.5	-10.1232\\
0.5	-10.1227\\
0.6	-10.2237\\
0.6	-10.2232\\
0.7	-10.3241\\
0.7	-10.3236\\
0.8	-10.3842\\
0.8	-10.385\\
0.9	-10.5843\\
0.9	-10.5846\\
};
\addlegendentry{D= 5}

\end{axis}
\end{tikzpicture}%
		\caption{Odd parity}
		\label{fig:energy_odd}
	\end{subfigure}
	\caption{Energies of the $3\times 4$ hexagonal lattice with $\kappa=1$, $U=2$ and $B=0$ at different values of $\mu$. Duplicate points correspond to $\chi=2D$ and $\chi=3D$.  See~\cite{Schneider2021,raw_data} for details.}
	\label{fig:energy_sectors}
\end{figure}
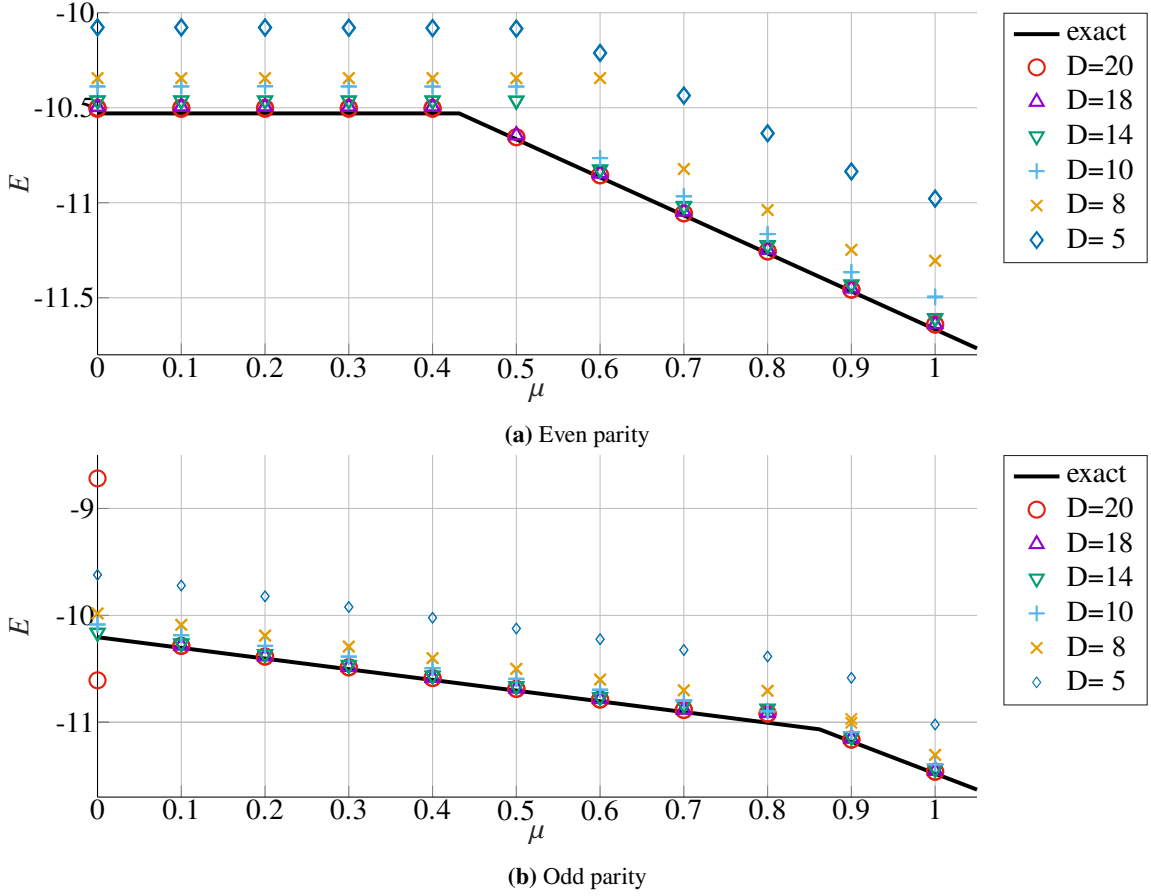
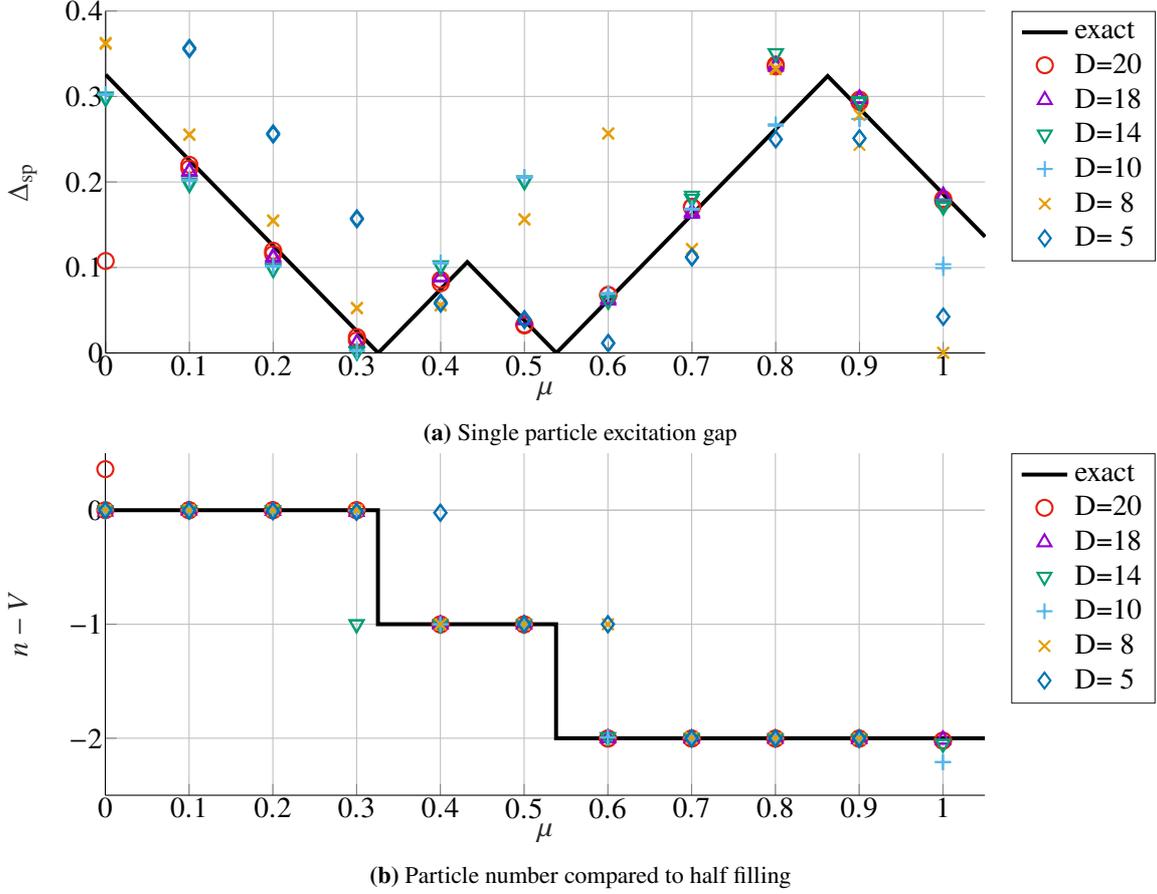
\begin{figure}[ht]
	\centering
	\begin{subfigure}[b]{\textwidth}
		\centering
%
%
\definecolor{mycolor1}{rgb}{0.89800,0.12200,0.05900}%
\definecolor{mycolor2}{rgb}{0.58000,0.00000,0.83100}%
\definecolor{mycolor3}{rgb}{0.00000,0.62000,0.45100}%
\definecolor{mycolor4}{rgb}{0.34100,0.71000,0.91000}%
\definecolor{mycolor5}{rgb}{0.89800,0.62000,0.00000}%
\definecolor{mycolor6}{rgb}{0.00000,0.45100,0.70200}%
\begin{tikzpicture}

\begin{axis}[%
width=0.765\textwidth,
height=\figHeight,
at={(0\textwidth,0\textwidth)},
scale only axis,
xmin=0,
xmax=1.05,
xlabel style={font=\color{white!15!black}},
xlabel={$\mu$},
ymin=0,
ymax=0.4,
ylabel style={font=\color{white!15!black}},
ylabel={$\Delta{}_{\text{sp}}$},
axis background/.style={fill=white},
axis x line*=bottom,
axis y line*=left,
xmajorgrids,
ymajorgrids,
legend style={at={(1.03,1)}, anchor=north west, legend cell align=left, align=left, draw=white!15!black},
ylabel style={{rotate=-90}}, scaled ticks=false, tick label style={/pgf/number format/fixed},
extra y ticks={0.4},
extra y tick labels={\phantom{-11.5}}
]
\addplot [color=black, line width=1.5]
  table[row sep=crcr]{%
0	0.325482933580501\\
0.325482933580501	0\\
0.43181417308775	0.106331239507249\\
0.538145412595	0\\
0.86203046922805	0.323885056633049\\
1.05	0.135915525861099\\
};
\addlegendentry{exact}

\addplot [color=mycolor1, only marks, mark=o, mark options={solid, mycolor1, scale=1.5, thick}]
  table[row sep=crcr]{%
0	0.107599747113699\\
0	1.7873886305375\\
0.1	0.215746490961401\\
0.1	0.220044618301301\\
1	0.1775648893093\\
1	0.180150280691201\\
0.2	0.1155791334991\\
0.2	0.119619243445699\\
0.3	0.0148715890363\\
0.3	0.0185675616637013\\
0.4	0.0855479923465001\\
0.4	0.0815687688994995\\
0.5	0.0322714440164003\\
0.5	0.0331903583272997\\
0.6	0.0680357574855996\\
0.6	0.0677379259003992\\
0.7	0.170508876427299\\
0.7	0.1709361246717\\
0.8	0.3370202674509\\
0.8	0.335129522998701\\
0.9	0.293542835554399\\
0.9	0.2963701034263\\
};
\addlegendentry{D=20}

\addplot [color=mycolor2, only marks, mark=triangle, mark options={solid, mycolor2, scale=1.5, thick}]
  table[row sep=crcr]{%
0	1.9419733045755\\
0	1.3420366379012\\
0.1	0.2119052329351\\
0.1	0.212871751757099\\
1	0.1828704468114\\
1	0.1842168287065\\
0.2	0.1108296319942\\
0.2	0.1126698817055\\
0.3	0.0119976264209001\\
0.3	0.0119392614774991\\
0.4	0.0891291529373994\\
0.4	0.088441202933101\\
0.5	0.0388366814185996\\
0.5	0.0382281513241001\\
0.6	0.0613188478979989\\
0.6	0.0634804231615007\\
0.7	0.162756810844201\\
0.7	0.1649404577313\\
0.8	0.334840845972701\\
0.8	0.336912800336\\
0.9	0.2978824174868\\
0.9	0.2970588858753\\
};
\addlegendentry{D=18}

\addplot [color=mycolor3, only marks, mark=triangle, mark options={solid, rotate=180, mycolor3, scale=1.5, thick}]
  table[row sep=crcr]{%
0	0.299944802738899\\
0	0.297860219459899\\
0.1	0.199483169332701\\
0.1	0.197281179477599\\
1	0.170952719447099\\
1	0.174387389478699\\
0.2	0.0990422397529009\\
0.2	0.0977877506788012\\
0.3	0.00122378661600031\\
0.3	0.00257677160169933\\
0.4	0.101315363233001\\
0.4	0.102595254482299\\
0.5	0.200459112276501\\
0.5	0.201971378172701\\
0.6	0.0611181819158997\\
0.6	0.0607542331005\\
0.7	0.180449043086298\\
0.7	0.183910043737601\\
0.8	0.3504604639981\\
0.8	0.3506440007429\\
0.9	0.293257714485101\\
0.9	0.2947481179588\\
};
\addlegendentry{D=14}

\addplot [color=mycolor4, only marks, mark=+, mark options={solid, mycolor4, scale=1.5, thick}]
  table[row sep=crcr]{%
0	0.302831909674399\\
0	0.301837043266399\\
0.1	0.202560537593\\
0.1	0.2016229392545\\
1	0.103786812230899\\
1	0.0991361936709989\\
0.2	0.102053525203399\\
0.2	0.100876796154999\\
0.3	0.00264068345079949\\
0.3	0.00185204075350143\\
0.4	0.104637164818101\\
0.4	0.105746716080899\\
0.5	0.2048164980145\\
0.5	0.205965372342101\\
0.6	0.0701263987189993\\
0.6	0.0687259431015992\\
0.7	0.168801760258701\\
0.7	0.167638494118199\\
0.8	0.267436543327101\\
0.8	0.2657589952197\\
0.9	0.273669603364299\\
0.9	0.273635380880901\\
};
\addlegendentry{D=10}

\addplot [color=mycolor5, only marks, mark=x, mark options={solid, mycolor5, scale=1.5, thick}]
  table[row sep=crcr]{%
0	0.3629066281981\\
0	0.3612612759339\\
0.1	0.2557982238301\\
0.1	0.254840173763501\\
1	0.000308374928501109\\
1	0.000204687834399309\\
0.2	0.1552499393303\\
0.2	0.1543367992347\\
0.3	0.0529967193711993\\
0.3	0.0520293872783011\\
0.4	0.0553349967873995\\
0.4	0.0562838037738995\\
0.5	0.1558746733634\\
0.5	0.1566946645616\\
0.6	0.256346965656901\\
0.6	0.256998781085301\\
0.7	0.1218361648016\\
0.7	0.1211739738321\\
0.8	0.333375281750401\\
0.8	0.3316720698294\\
0.9	0.243286889236801\\
0.9	0.2781412388606\\
};
\addlegendentry{D= 8}

\addplot [color=mycolor6, only marks, mark=diamond, mark options={solid, mycolor6, scale=1.5, thick}]
  table[row sep=crcr]{%
0	0.455654355512099\\
0	0.457222460760599\\
0.1	0.3551792867472\\
0.1	0.3568866056128\\
1	0.0424659339289999\\
1	0.0423908695109994\\
0.2	0.2552680580798\\
0.2	0.257042160020301\\
0.3	0.156144540023901\\
0.3	0.1577031710436\\
0.4	0.0575211356125997\\
0.4	0.0590841238161008\\
0.5	0.0397832249780006\\
0.5	0.0383507932318992\\
0.6	0.0120734001697986\\
0.6	0.0110873407253003\\
0.7	0.1114980491288\\
0.7	0.112433295700098\\
0.8	0.2501603410158\\
0.8	0.249744983301801\\
0.9	0.251042420712599\\
0.9	0.251021046758201\\
};
\addlegendentry{D= 5}

\end{axis}
\end{tikzpicture}%
		\caption{Single particle excitation gap}
		\label{fig:energy_gap}
	\end{subfigure}
	\begin{subfigure}[b]{\textwidth}
		\centering
%
%
\definecolor{mycolor1}{rgb}{0.89800,0.12200,0.05900}%
\definecolor{mycolor2}{rgb}{0.58000,0.00000,0.83100}%
\definecolor{mycolor3}{rgb}{0.00000,0.62000,0.45100}%
\definecolor{mycolor4}{rgb}{0.34100,0.71000,0.91000}%
\definecolor{mycolor5}{rgb}{0.89800,0.62000,0.00000}%
\definecolor{mycolor6}{rgb}{0.00000,0.45100,0.70200}%
\begin{tikzpicture}

\begin{axis}[%
width=0.765\textwidth,
height=\figHeight,
at={(0\textwidth,0\textwidth)},
scale only axis,
xmin=0,
xmax=1.05,
xlabel style={font=\color{white!15!black}},
xlabel={$\mu$},
ymin=-2.5,
ymax=0.5,
ylabel style={font=\color{white!15!black}},
ylabel={$\phantom{\Delta{}_{\text{sp}}}n - V\phantom{\Delta{}_{\text{sp}}}$},
axis background/.style={fill=white},
axis x line*=bottom,
axis y line*=left,
xmajorgrids,
ymajorgrids,
legend style={at={(1.03,1)}, anchor=north west, legend cell align=left, align=left, draw=white!15!black},
ylabel style={{rotate=-90}}, scaled ticks=false, tick label style={/pgf/number format/fixed},
extra y ticks={0},
extra y tick labels={\phantom{-11.5}}
]
\addplot [color=black, line width=1.5]
table[row sep=crcr]{%
	0	0\\
	0.325482933580501	0\\
	0.325482933580501	-1\\
	0.538145412595	-1\\
	0.538145412595	-2\\
	1.05	-2\\
};
\addlegendentry{exact}

\addplot [color=mycolor1, only marks, mark=o, mark options={solid, mycolor1, scale=1.5, thick}]
table[row sep=crcr]{%
	0	0.359905\\
	0	5.80681e-07\\
	0.1	-0.000310678\\
	0.1	-0.000109568\\
	1	-2.02243\\
	1	-2.01998\\
	0.2	-0.000640977\\
	0.2	-0.000212761\\
	0.3	-0.000946532\\
	0.3	-0.000327728\\
	0.4	-1.00201\\
	0.4	-1.00112\\
	0.5	-1.00223\\
	0.5	-1.0011\\
	0.6	-2.00324\\
	0.6	-2.00215\\
	0.7	-2.0031\\
	0.7	-2.00233\\
	0.8	-2.00324\\
	0.8	-2.00238\\
	0.9	-2.00424\\
	0.9	-2.00254\\
};
\addlegendentry{D=20}

\addplot [color=mycolor2, only marks, mark=triangle, mark options={solid, mycolor2, scale=1.5, thick}]
table[row sep=crcr]{%
	0	-0.0129274\\
	0	12.9672\\
	0.1	-0.00198665\\
	0.1	-0.00182546\\
	1	-2.01241\\
	1	-2.01131\\
	0.2	-0.00562777\\
	0.2	-0.00538667\\
	0.3	-0.015101\\
	0.3	-0.0148466\\
	0.4	-1.00163\\
	0.4	-1.00069\\
	0.5	-1.00206\\
	0.5	-1.00097\\
	0.6	-2.00086\\
	0.6	-1.99876\\
	0.7	-2.00316\\
	0.7	-2.00106\\
	0.8	-2.00369\\
	0.8	-2.00201\\
	0.9	-2.00675\\
	0.9	-2.00453\\
};
\addlegendentry{D=18}

\addplot [color=mycolor3, only marks, mark=triangle, mark options={solid, rotate=180, mycolor3, scale=1.5, thick}]
table[row sep=crcr]{%
	0	-5.68029e-08\\
	0	2.18148e-06\\
	0.1	-0.000117807\\
	0.1	-0.000109699\\
	1	-2.04888\\
	1	-2.04872\\
	0.2	-0.000386014\\
	0.2	-0.000327532\\
	0.3	-1.00098\\
	0.3	-1.00041\\
	0.4	-1.00097\\
	0.4	-1.00046\\
	0.5	-1.00093\\
	0.5	-1.00046\\
	0.6	-1.99123\\
	0.6	-1.99012\\
	0.7	-1.9991\\
	0.7	-1.99875\\
	0.8	-1.99965\\
	0.8	-1.99974\\
	0.9	-2.00713\\
	0.9	-2.00542\\
};
\addlegendentry{D=14}

\addplot [color=mycolor4, only marks, mark=+, mark options={solid, mycolor4, scale=1.5, thick}]
table[row sep=crcr]{%
	0	5.76782e-05\\
	0	4.11435e-05\\
	0.1	-0.00084529\\
	0.1	-0.00083784\\
	1	-2.20881\\
	1	-2.20983\\
	0.2	-0.00013473\\
	0.2	-8.05934e-05\\
	0.3	-0.00395075\\
	0.3	-0.00389784\\
	0.4	-1.00166\\
	0.4	-1.00015\\
	0.5	-1.00183\\
	0.5	-1.00016\\
	0.6	-1.99435\\
	0.6	-1.99281\\
	0.7	-1.99855\\
	0.7	-1.99714\\
	0.8	-2.00181\\
	0.8	-2.00005\\
	0.9	-2.00212\\
	0.9	-2.0001\\
};
\addlegendentry{D=10}

\addplot [color=mycolor5, only marks, mark=x, mark options={solid, mycolor5, scale=1.5, thick}]
table[row sep=crcr]{%
	0	6.6058e-09\\
	0	-4.33488e-06\\
	0.1	-2.10788e-05\\
	0.1	-1.35061e-05\\
	1	-3.00025\\
	1	-3.00019\\
	0.2	-4.23097e-05\\
	0.2	-4.41002e-05\\
	0.3	-6.35675e-05\\
	0.3	-5.4639e-05\\
	0.4	-1.0011\\
	0.4	-1.00004\\
	0.5	-1.00114\\
	0.5	-1.00005\\
	0.6	-1.00118\\
	0.6	-1.00004\\
	0.7	-1.99687\\
	0.7	-1.99638\\
	0.8	-2.00078\\
	0.8	-1.99998\\
	0.9	-2.00404\\
	0.9	-2.00365\\
};
\addlegendentry{D= 8}

\addplot [color=mycolor6, only marks, mark=diamond, mark options={solid, mycolor6, scale=1.5, thick}]
table[row sep=crcr]{%
	0	-6.5142e-05\\
	0	-1.00336e-05\\
	0.1	-0.00403624\\
	0.1	-0.0039041\\
	1	-2.99995\\
	1	-2.99999\\
	0.2	-0.00845319\\
	0.2	-0.00808781\\
	0.3	-0.0143985\\
	0.3	-0.0135976\\
	0.4	-0.022745\\
	0.4	-0.0223026\\
	0.5	-1.00005\\
	0.5	-0.999988\\
	0.6	-1.00006\\
	0.6	-0.999988\\
	0.7	-2.00092\\
	0.7	-2.00002\\
	0.8	-2.00092\\
	0.8	-2.00001\\
	0.9	-2.00095\\
	0.9	-2.00002\\
};
\addlegendentry{D= 5}

\end{axis}
\end{tikzpicture}%
		\caption{Particle number compared to half filling}
		\label{fig:particle_number}
	\end{subfigure}
	\caption{Single particle gap and particle number of the $3\times 4$ hexagonal lattice with $\kappa=1$, $U=2$ and $B=0$ at different values of $\mu$. Duplicate points correspond to $\chi=2D$ and $\chi=3D$. See~\cite{Schneider2021,raw_data} for details and calculated values.}
\end{figure}

Only in cases where the ground state in one sector is degenerate or nearly degenerate, we see significant deviations from the exact results. This is the case in the odd parity sector at $\mu=0$ and close to the kinks, where the particle number of the ground state changes. Otherwise we obtain reliable results, also for values of the chemical potential where Monte Carlo methods suffer from a very severe sign problem \cite{Schneider2021}.

We are able to reproduce the expected step function of the particle number as shown in \cref{fig:particle_number}. A similar study has been done using the tensor renormalization group approach for the Hubbard model on a square lattice~\cite{TRG_Akiyama:2021}. We compare the energies of the two independent sectors to decide which sector corresponds to the ground state. If the system is close to a phase transition the single particle gap becomes small, which can lead to a wrong choice due to the errors of the energy calculations. This can be seen in \cref{fig:particle_number} close to the steps, where occasionally the particle number of the wrong phase got chosen. We see that the particle number becomes correct if we increase the bond dimension and therefore the precision of the calculated energies. An alternative approach to find the phase transitions was used in~\cite{Schwinger_Banuls:2016}. The energy scales linearly in the chemical potential within one phase. Determining the intersections with a linear fit of several data points with varying $\mu$ can help to find the phase transitions more accurately. We plan to use this method in the future for a better determination of the transition points.

\begin{figure}[ht]
	\centering
	\begin{subfigure}[b]{\textwidth}
		\centering
%
%
\definecolor{mycolor1}{rgb}{0.58000,0.00000,0.82700}%
\definecolor{mycolor2}{rgb}{0.00000,0.62000,0.45100}%
\definecolor{mycolor3}{rgb}{0.33700,0.70600,0.91400}%
\definecolor{mycolor4}{rgb}{0.90200,0.62400,0.00000}%
\definecolor{mycolor5}{rgb}{0.89800,0.11800,0.06300}%
\begin{tikzpicture}

\begin{axis}[%
width=0.664\textwidth,
height=\figHeight,
at={(0\textwidth,0\textwidth)},
scale only axis,
xmin=0,
xmax=0.07,
xlabel style={font=\color{white!15!black}},
xlabel={$D^{-2}$},
ymin=-700,
ymax=-610,
ylabel style={font=\color{white!15!black}},
ylabel={$E$},
axis background/.style={fill=white},
axis x line*=bottom,
axis y line*=left,
xmajorgrids,
ymajorgrids,
legend style={at={(1.03,1)}, anchor=north west, legend cell align=left, align=left, draw=white!15!black},
ylabel style={{rotate=-90}}, scaled ticks=false, tick label style={/pgf/number format/fixed}
]
\addplot [color=mycolor1, only marks, mark=triangle, mark options={solid, mycolor1, scale=1.5, thick}]
  table[row sep=crcr]{%
0.01	-682.034446135984\\
0.01	-681.922920335153\\
0.0625	-625.966975417992\\
0.0625	-625.927338226215\\
0.04	-644.108695804626\\
0.04	-644.13285363416\\
0.0277777777777778	-667.117126757106\\
0.0277777777777778	-667.075820000383\\
0.0204081632653061	-670.557199017034\\
0.0204081632653061	-670.616585764682\\
0.015625	-674.984907136752\\
0.0123456790123457	-679.612999435687\\
};
\addlegendentry{even, simulation}

\addplot [color=mycolor2, only marks, mark=triangle, mark options={solid, rotate=180, mycolor2, scale=1.5, thick}]
  table[row sep=crcr]{%
0.01	-681.728\\
0.01	-681.605\\
0.0625	-626.581\\
0.0625	-626.539\\
0.04	-642.658\\
0.04	-642.686\\
0.0277777777777778	-665.158\\
0.0277777777777778	-665.123\\
0.0204081632653061	-670.206\\
0.0204081632653061	-670.261\\
0.015625	-674.279\\
0.0123456790123457	-679.045\\
};
\addlegendentry{odd, simulation}

\addplot [color=mycolor3, line width=1.5]
  table[row sep=crcr]{%
0	-693.173\\
0.07	-616.2241\\
};
\addlegendentry{even, fit}

\addplot [color=mycolor4, line width=1.5]
  table[row sep=crcr]{%
0	-692.096\\
0.07	-616.0956\\
};
\addlegendentry{odd, fit}

\addplot [color=black, line width=1.5]
  table[row sep=crcr]{%
0	-692.275\\
0.07	-692.275\\
};
\addlegendentry{even, exact}

\addplot [color=mycolor5, dotted, line width=1.5]
  table[row sep=crcr]{%
0	-692.247\\
0.07	-692.247\\
};
\addlegendentry{odd, exact}

\end{axis}
\end{tikzpicture}%
		\caption{$U=0$}
		\label{fig:energy_large_U0}
	\end{subfigure}
	\begin{subfigure}[b]{\textwidth}
		\centering
%
%
\definecolor{mycolor1}{rgb}{0.58000,0.00000,0.82700}%
\definecolor{mycolor2}{rgb}{0.00000,0.62000,0.45100}%
\definecolor{mycolor3}{rgb}{0.33700,0.70600,0.91400}%
\definecolor{mycolor4}{rgb}{0.90200,0.62400,0.00000}%
\begin{tikzpicture}

\begin{axis}[%
width=0.664\textwidth,
height=\figHeight,
at={(0\textwidth,0\textwidth)},
scale only axis,
xmin=0,
xmax=0.07,
xlabel style={font=\color{white!15!black}},
xlabel={$D^{-2}$},
ymin=-490,
ymax=-420,
ylabel style={font=\color{white!15!black}},
ylabel={$E$},
axis background/.style={fill=white},
axis x line*=bottom,
axis y line*=left,
xmajorgrids,
ymajorgrids,
legend style={at={(1.03,1)}, anchor=north west, legend cell align=left, align=left, draw=white!15!black},
ylabel style={{rotate=-90}}, scaled ticks=false, tick label style={/pgf/number format/fixed}
]
\addplot [color=mycolor1, only marks, mark=triangle, mark options={solid, mycolor1, scale=1.5, thick}]
  table[row sep=crcr]{%
0.01	-476.99338406838\\
0.0625	-432.403492008337\\
0.0625	-432.389527839572\\
0.04	-446.414876604981\\
0.04	-446.408939451364\\
0.0277777777777778	-463.388213240722\\
0.0277777777777778	-463.531229886423\\
0.0204081632653061	-465.623475340259\\
0.0204081632653061	-465.745373526401\\
0.015625	-469.316886730929\\
0.015625	-469.363692262832\\
0.0123456790123457	-473.041765718215\\
};
\addlegendentry{even, simulation}

\addplot [color=mycolor2, only marks, mark=triangle, mark options={solid, rotate=180, mycolor2, scale=1.5, thick}]
  table[row sep=crcr]{%
0.01	-476.384\\
0.0625	-430.717\\
0.0625	-430.701\\
0.04	-445.73\\
0.04	-445.728\\
0.0277777777777778	-462.191\\
0.0277777777777778	-462.317\\
0.0204081632653061	-465.851\\
0.0204081632653061	-465.969\\
0.015625	-468.917\\
0.015625	-468.964\\
0.0123456790123457	-474.022\\
};
\addlegendentry{odd, simulation}

\addplot [color=mycolor3, line width=1.5]
  table[row sep=crcr]{%
0	-483.46\\
0.07	-424.91158\\
};
\addlegendentry{even, fit}

\addplot [color=mycolor4, line width=1.5]
  table[row sep=crcr]{%
0	-483.835\\
0.07	-422.99275\\
};
\addlegendentry{odd, fit}

\end{axis}
\end{tikzpicture}%
		\caption{$U=2$}
		\label{fig:energy_large_U2}
	\end{subfigure}
	\caption{Energies with finite chemical potential ($\kappa=1$, $\mu=0.5$, $B=0$) for the $30\times 15$-lattice against the inverse squared bond dimension. Duplicate points correspond to $\chi=2D$ and $\chi=3D$. See~\cite{Schneider2021,raw_data} for details.}
	\label{fig:large_volume}
\end{figure}
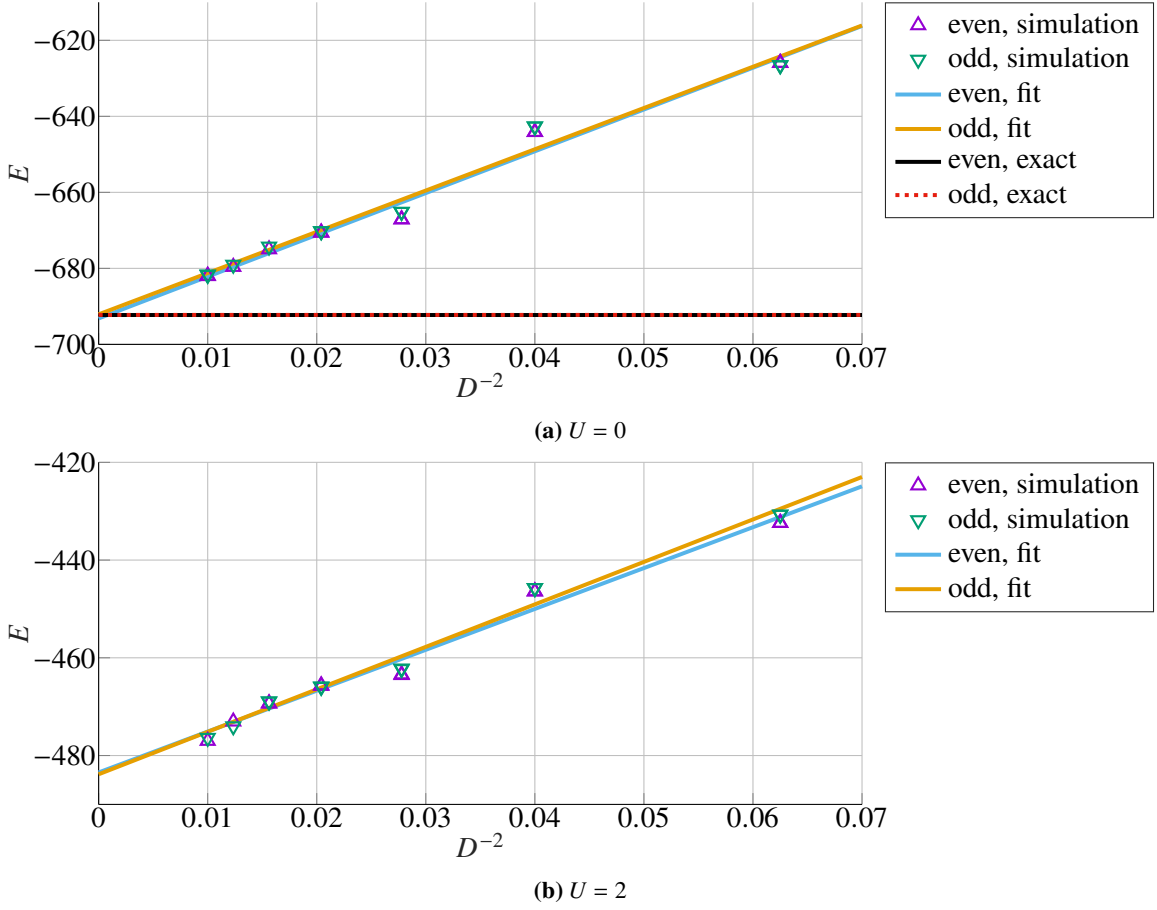

After this proof of principle we simulate a larger $30 \times 15$ lattice to show the scalability of the algorithm. In \cref{fig:large_volume} we present the energies for a finite chemical potential in the non-interacting limit and for $U=2$. We find empirically that the error of the energy scales as $D^{-2}$ in the bond dimension. This allows us to extrapolate to the infinite bond dimension limit. The extrapolated results for $U=0$ are compatible with the energies from an exact diagonalization of the hopping matrix: $E_\text{even} = -693.2(17)$ and $E_\text{odd} = -692.1(18)$; the exact energies are $E_\text{exact} \approx -692.3$ for both sectors on our level of precision. In the interacting case no exact results are available, which demonstrates that we can use our method to explore new, unknown areas of the phase space. Our final results for the energies of the system in~\cref{fig:energy_large_U2} are $E_\text{even} = {-483.5(14)}$ and $E_\text{odd} = {-483.8(12)}$.

\section{Conclusion}
We studied the Hubbard model using a fermionic PEPS algorithm. The method was successfully applied to cases with a chemical potential, where Monte Carlo methods fail due to the sign problem. Using an additional parity link on one of the PEPS tensors allowed us to study physical systems in both parity sectors, which is a great novelty.

To calculate expectation values, we used the boundary matrix product state method with a truncated bond dimension $\chi$. Our analyses showed that a linear scaling of $\chi$ in the PEPS bond dimension $D$ was sufficient and the popularly used~\cite{tensor_fermions,Lubasch_algorithms} scaling $\chi \propto D^2$ was too conservative for this particular system. A smaller $\chi$ allowed us to increase $D$ and gain an overall higher precision with limited computational resources.

The ground state energies were calculated to high accuracy for a small system size as well as for a large system with vanishing onsite interaction. In the latter case an accuracy of the order of $10^{-3}$ was achieved with extrapolations in the squared inverse bond dimension. We also simulated a system at large volume with onsite interaction and chemical potential. This opens the path to exploring the phase diagram of the Hubbard model also when a chemical potential is added, an enterprise we want to follow in the future.

\acknowledgments
We thank Stefan Kühn, Román Orús and Mari Carmen Bañuls for helpful discussions and remarks. We acknowledge the Computer Center at DESY Zeuthen for the compute time. This work was funded, in part, through financial support by the Helmholtz Einstein International Berlin Research School in Data Science (HEIBRiDS), by the Deutsche Forschungsgemeinschaft (DFG, German Research Foundation) and by the NSFC through the funds provided to the Sino-German Collaborative Research Center CRC 110 “Symmetries and the Emergence of Structure in QCD” (DFG Project-ID 196253076 - TRR 110, NSFC Grant No. 12070131001).

%

\bibliographystyle{JHEP}
\bibliography{bibliography}

\end{document}